\documentclass[aps,prx,preprint,superscriptaddress,notitlepage]{revtex4-2}

\usepackage{graphicx}
\usepackage{dcolumn}
\usepackage{bm}
\usepackage[version=4]{mhchem}
\usepackage{blindtext}
\usepackage{amsmath,amssymb}
\usepackage{upgreek}

\usepackage{hyperref}

\begin{document}

\title{Direct link between disorder, mobility and magnetoresistance in topological semimetals}
\author{Jocienne N. Nelson}
\affiliation{National Renewable Energy Laboratory, Golden, Colorado 80401, USA}
\author{Anthony D. Rice}
\affiliation{National Renewable Energy Laboratory, Golden, Colorado 80401, USA}
\author{Chase Brooks}
\affiliation{Department of Physics, University of Colorado, Boulder, CO 80309}
\author{Ian A. Leahy}
\affiliation{Department of Physics, University of Colorado, Boulder, CO 80309}
\author{Glenn Teeter}
\affiliation{National Renewable Energy Laboratory, Golden, Colorado 80401, USA}
\author{Mark Van Schilfgaarde}
\affiliation{National Renewable Energy Laboratory, Golden, Colorado 80401, USA}
\author{Stephan Lany}
\affiliation{National Renewable Energy Laboratory, Golden, Colorado 80401, USA}
\author{Brian Fluegel}
\affiliation{National Renewable Energy Laboratory, Golden, Colorado 80401, USA}
\author{Minhyea Lee}
\affiliation{Department of Physics, University of Colorado, Boulder, CO 80309}
\author{Kirstin Alberi}
\affiliation{National Renewable Energy Laboratory, Golden, Colorado 80401, USA}
\email[]{To whom all correspondence should be addressed. Kirstin.Alberi@nrel.gov}

\begin{abstract}
The extent to which disorder influences the properties of topological semimetals remains an open question and is relevant to both the understanding of topological states and the use of topological materials in practical applications. Here, we achieve unmatched and systematic control of point defect concentrations in the prototypical Dirac semimetal Cd$_3$As$_2$ to gain important insight into the role of disorder on electron transport behavior. We find that arsenic vacancies introduce localized states near the Fermi level and strongly influence the electron mobility. Reducing arsenic vacancies by changing the As/Cd flux ratio used during deposition results in an increase in the magnetoresistance from 200\% – 1000\% and an increase in mobility from 5000 – 18,000 cm$^2$/Vs. However, the degree of linear magnetoresistance, which has previously been linked to disorder, is found here to correlate inversely with measures of disorder, including disorder potential and disorder correlation lengths. This finding yields important new information in the quest to identify the origin of linear magnetoresistance in a wider range of materials.

\end{abstract}

\maketitle

\section{Introduction}

Three dimensional topological semimetals (TSMs) exhibit linear and gapless bulk band dispersions along with topologically protected surface states arising from the band inversion \cite{schoop_chemical_2018}. In particular, topological protection against carrier backscattering is thought to help support high electron mobilities that have been measured up to 10$^7$ cm$^2$/Vs in the Dirac semimetal Cd$_3$As$_2$ and 10$^5$ - 10$^6$ cm$^2$/Vs in the Weyl semimetal family of (Ta,Nb)(As,P) \cite{LiangNmat2015,Singh2020}. These materials also typically feature Fermi levels near the band touching nodes ($\sim$100 meV) coupled with large Fermi velocities, which can only occur in systems with non-parabolic bands. Together, these distinct aspects lead to extremely large, non-saturating linear magnetoresistance (LMR) that occurs at experimentally-accessible magnetic fields. Most studies agree that disorder is the basis for non-saturating LMR \cite{LeahyPNAS2018,Parish2005,Kisslinger2017} but do not provide a clear picture of the exact mechanisms that influence it or how to tune it. To address this gap, we systematically vary the point defect populations in the prototypical Dirac semimetal Cd$_3$As$_2$ to directly obtain information about how disorder influences both mobility and LMR in TSMs.

The role of disorder is particularly important to understand, as thin film TSMs have recently become available \cite{Schumann2016,BedoyaPinto2020,Asaba2018}  and will ultimately be integrated into low power electronic and optoelectronic devices. Real TSM crystals and epilayers naturally contain disorder introduced by point and extended defects. The extent to which they influence transport behavior when the topological properties are regarded to be robust against small scales of chemical and structural variations remains an open question. Theoretical studies generally agree that short range disorder \cite{Nandkishore2014, Pixley2016,Holder2017} (i.e. from neutral impurities and defects) can lead to a nonzero density of states at the Dirac point and a finite mean-free path, while long range disorder \cite{SkinnerPRB2014} (i.e. from charged impurities and defects) can additionally introduce electron and hole “puddles” that also dictate the disorder potential and length scale. Experimentally, few studies have revealed the true impact of disorder on transport. Qualitative observation of conductivity fluctuations\cite{Schumann2017} and quantitative detection of disorder potentials on the order of a few 10s of meV \cite{jeon_landau_2014} have been made in single samples of the Dirac semimetal Cd$_3$As$_2$, and similar disorder potentials and mean free paths on the order of a few 10s of nm have been extracted from MR measurements in Weyl semimetals TaP and NbP\cite{LeahyPNAS2018}. However, the absence of systematic variation of specific instances of disorder have so far prevented a deeper understanding of the true role of disorder on electron transport and ways in which it can be controlled to manipulate TSM properties.

In this work, we quantitatively extract disorder potentials and electron mean free paths as a function of continuously varying native point defect concentrations in the prototypical Dirac semimetal Cd$_3$As$_2$. Specifically, we use epitaxial growth with separate Cd and As sources to modify the chemical potential during growth and methodically tune the native defect populations to a degree that is not available in bulk synthesis. We show that native point defects, which are likely As vacancies, introduce disorder potentials on the order of 15-25 meV and limit the electron mean free path to 20-80 nm. Importantly, the magnitude of LMR correlates inversely with the density of As vacancies and is strongly determined by the electron mobility, providing new insights into the origins of large LMR.

\section{Methods}

Cd$_3$As$_2$(112) thin films were synthesized at 115$^\circ$C on GaAs(111) substrates using light-assisted molecular beam epitaxy, as previously reported in Ref \cite{Rice2019}. The ratio of the As/Cd fluxes were intentionally varied from 0.17-2.37. We note that this level of variation in the flux ratio does not influence the stoichiometry of the Cd$_3$As$_2$ epilayers at the atomic percent level, as demonstrated by x-ray photoemission spectroscopy measurements (Fig. \ref{fig:FigXPS}) but instead influences the concentration of native point defects at typical electronic doping levels. Hall bars were fabricated with standard photolithography, wet chemical etching and electroplated Au contacts. Magnetotransport measurements were performed in a Physical Property Measurement System from Quantum Design. We also present a Density Functional Theory (DFT) analysis of the intrinsic defects for the 80-atom ground state structure with spin-orbit coupling included. Total energies and electronic structures were calculated using the pseudopotential momentum-space formalism, the Projector Augmented Wave (PAW) method, a 4 $\times$ 4 $\times$ 4 $\Gamma$-centered k-mesh, and the Strongly Constrained and Appropriately Normed (SCAN) metageneralized gradient approximation functional \cite{Sun2015}, as implemented in the VASP code \cite{Kresse1999}.More information on the synthesis, magnetotransport measurements and calculations are available in the supplementary information and Ref. \cite{Rice2019}.

\section{Results}

\subsection{Magnetotransport measurements of Cd$_3$As$_2$}

\begin{figure*} 
	\includegraphics[width=1\linewidth]{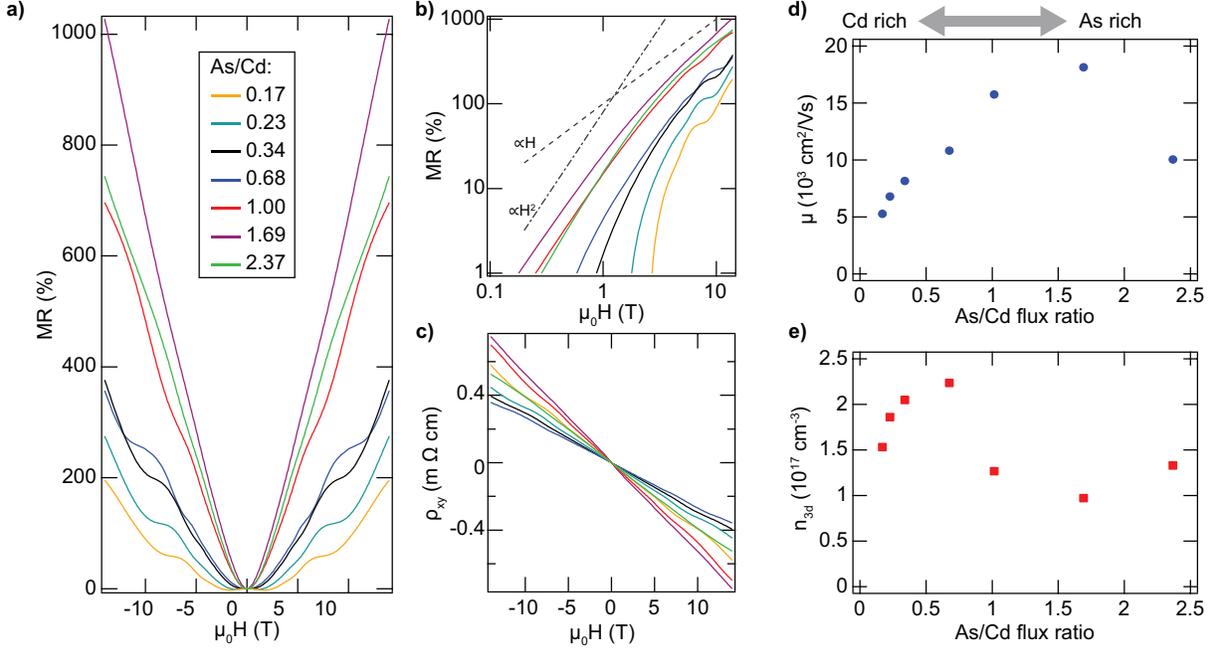}
	\caption{\label{fig:Fig1} Electrical transport measurements performed at 2 K. a) Magnetoresistance \% as a function of applied magnetic field  H $\parallel\hat{z}$, current I $\parallel\hat{x}$ for Cd$_3$As$_2$ samples grown with different As/Cd ratios but otherwise under identical conditions. b) Log-log plot illustrating that all samples crossover to a linear field dependence at a saturation field H$_\mathrm{S}$ that depends on the As/Cd ratio. c) Hall resistivity as a function of magnetic field for the same samples. d) Mobility $\mu$ and e) electron concentration n$_\mathrm{3D}$ as a function of As/Cd ratio.}
\end{figure*} 

Figure \ref{fig:Fig1} presents results demonstrating that the As/Cd flux ratio used during growth significantly impacts the MR properties. All measurements were performed at 2 K. Figure \ref{fig:Fig1}(a) shows the MR \% defined as MR\%=100\% $\times \frac{(\rho_{xx}(\mu_0 H)-\rho_{xx}(\mu_0 H=0))}{\rho_{xx}(\mu_0 H=0))}$, where $\rho_{xx}$ is the longitudinal resistivity and the magnetic field H $\parallel\hat{z}$, current I $\parallel\hat{x}$. The MR at $\mu_0$H=14 T ranges from 196\% in the sample grown under the lowest As/Cd flux ratio (As/Cd = 0.17, yellow) to 1028\% in the sample grown with the nearly highest As/Cd flux ratio (As/Cd =1.69, purple). This later value is large compared to 300-500 \% values measured in other Cd$_3$As$_2$ thin films \cite{Shoron2021,Nakazawa2019}, and the total MR\% variation across the samples measured is greater than 5x. Furthermore, as shown in the log-log plot in Fig. \ref{fig:Fig1}b) all samples display LMR with an onset field that depends on the As/Cd flux ratio. Figure \ref{fig:Fig1}(c) shows the Hall resistivity ($\rho_{xy}$) as a function of applied field. The $\rho_{xy}$ of all samples has a negative slope, suggesting that the high field MR is dominated by a single electronlike carrier. This behavior is consistent with measurements in bulk crystals and is expected in Cd$_3$As$_2$ when the Fermi level is pinned above the Dirac node \cite{LiangNmat2015, HePRL2014}. The presence of Shubnikov-de Haas (SdH) oscillations demonstrates that all samples are of high quality and allows us to extract material parameters by fitting to a standard Lifshitz-Kosevich formula (described in further detail in the supplemental information). Both the fitting and additional Fourier transform analysis (Fig. \ref{fig:Figrvst}d)) indicate that there is primarily one SdH oscillation frequency consistent with a single Dirac band crossing the Fermi level. Figures \ref{fig:Fig1}(d,e) show the mobility ($\mu$) and carrier concentration, n$_\mathrm{3d}$, extracted from low field ($\pm$0.1 T) Hall measurements as a function of As/Cd flux ratio.

The transport behavior evolves as a function of the As/Cd flux ratio used during growth. Samples synthesized with generally higher Cd fluxes (As/Cd $\lesssim$ 1) tend to have lower MR \% and smaller mobilities. At the other end of the spectrum, samples synthesized with higher As fluxes (As/Cd $\gtrsim$ 1) exhibit higher mobilities and MR\% that have largely saturated by the point where As/Cd=1.69. The decrease in mobility in the As/Cd=2.37 sample is likely due to the fact that this is at the bound of the accessible growth phase space where excess As begins to accumulate on the sample surface.

\begin{figure*} 
	\includegraphics[width=1\linewidth]{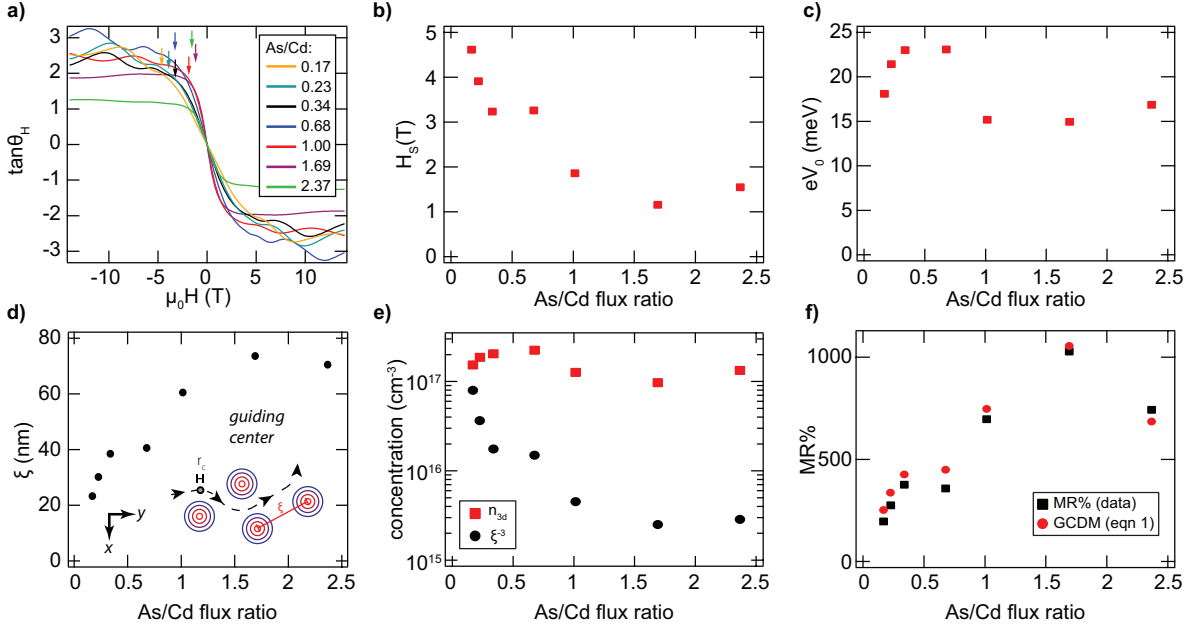}
	\caption{\label{fig:Fig2} a) $\tan(\theta_\mathrm{H}) = \frac{\rho_{xy}}{\rho_{xx}}$ as a function of magnetic field for samples grown under different As/Cd ratios. b) The field saturation of $\tan(\theta_\mathrm{H})$ which occurs at field $\mathrm{H_S}$ when MR\% or equivalently $\rho_{xx}$ becomes linear as $\rho_{xy}$ is linear over the whole field range. c) $eV_0$, where $V_0$ is the disorder potential as a function of As/Cd flux ratio. d) The disorder length scale $\xi$ determined by the radius of the cyclotron radius at the saturation field H$_\mathrm{S}$. The inset shows a schematic of the guiding center diffusive motion (GCDM), adapted from Ref. \cite{SongPRB2015}. e) A comparison between the carrier concentration n$_\mathrm{3d}$ (red squares), and the density of point defects suggested by the disorder length scale $\xi$ (black circles). f) MR\% as a function of the As/Cd ratio compared with the predicted MR\% from the GCDM \cite{SongPRB2015}. All measurements were performed at 2K.}
\end{figure*}

In Fig. \ref{fig:Fig2}, we turn to analysis of the magnetotransport in order to quantify the role of disorder in this system, which we will show is key to explaining the changes in MR\%, mobility and carrier concentration shown in Fig. \ref{fig:Fig1}. Figure \ref{fig:Fig2}(a) shows the field dependence of the tangent of the Hall angle $\tan(\theta_\mathrm{H}) = \frac{\rho_{xy}}{\rho_{xx}}$, measured at 2K. We observe a large $\tan(\theta_\mathrm{H})$ that saturates at a large value for all samples. Similar behavior of the tangent of the Hall angle has previously been used to determine the origins of non-saturating LMR in TaP and NbP \cite{LeahyPNAS2018}. In these materials, this characteristic field saturation of $\tan(\theta_\mathrm{H})$  was found to be explained by the guiding center diffusion model (GCDM) for linear MR \cite{SongPRB2015}.

The GCDM model does not require multiband compensation or significant disorder to explain the non saturating LMR and is the most appropriate framework for Cd$_3$As$_2$. A schematic of the GCDM principles is shown in the inset of Fig. \ref{fig:Fig2}(d). Trajectories of charge carriers are altered by a smoothly varying potential energy landscape originating from disorder with typical strength, $V_0$, and correlation length, $\xi$. Applying a magnetic field of sufficient strength to reduce the cyclotron radius, $r_c$, below $\xi$ forces the carrier’s guiding center of motion to diffuse around the potential barriers, resulting in LMR. In the case of Cd$_3$As$_2$, the relatively high Fermi velocities ($\sim$10$^6$ cm/s) \cite{LiangNmat2015} and small magnitudes of the Fermi level above the Dirac point (0.1-0.2 eV) suggests that $r_c$ can reach 10-100 nm at modest magnetic fields ($\sim$10 T), permitting us to probe disorder on this length scale. This combination of large Fermi velocity and low Fermi level is common to many other topological materials with non-parabolic bands, highlighting the large scope of applicability of our findings. Analysis of the $\tan(\theta_\mathrm{H})$ field dependence allows us to quantitatively determine when the MR becomes linear: because $\rho_{xy}$ is linear over the whole field range studied, $\tan(\theta_\mathrm{H})$ is field independent once $\rho_{xx}$ becomes linear. We extract the saturation field H$_\mathrm{S}$ (indicated by arrows in Fig. \ref{fig:Fig2}(a)) by fitting $\tan(\theta_\mathrm{H})$ to the empirical equation $\tan(\theta_\mathrm{H}) = A \tanh(\mathrm{H}/\mathrm{H}_\mathrm{S})$, where A is a constant. The extracted H$_\mathrm{S}$, shown in Fig. \ref{fig:Fig2}(b), decreases as the As/Cd flux ratio is increased and also is correlated with increased mobility.

From the GCDM framework, we extract the disorder strength, $V_0$ (Fig. \ref{fig:Fig2}(c)), which is predicted to be related to the tangent of the Hall angle by $\tan(\theta_\mathrm{H}) = \frac{2}{\sqrt{27\pi}}\left(\frac{E_F}{e V_0}\right)^{3/2}$  in the regime where $\tan(\theta_\mathrm{H})$ is field independent (i.e. H$>$H$_\mathrm{S}$)\cite{SongPRB2015}. Here the Fermi level $E_\mathrm{F}$, extracted via Lifshitz-Kosevich fitting of the SdH oscillations (supplementary note 3), is defined relative to the Dirac point. This demonstrates the $eV_0$ decreases with increasing As/Cd flux ratio. The values of $eV_0$ determined in our Cd$_3$As$_2$ epilayers are similar to those found in bulk NbP and TaP Weyl semimetal crystals (7 meV and 10 meV, respectively) \cite{LeahyPNAS2018}, providing additional evidence that the disorder potentials in topological semimetals are low compared to the thermal energy of electrons that extend up to even modest temperatures. The disorder correlation length can be extracted from the cyclotron radius at the magnetic field $\mathrm{H_S}$ where MR\% becomes linear: $\xi \approx \frac{m_e v_F}{e \mu_0 H_\mathrm{S}}$. In Fig. \ref{fig:Fig2}(d) we show $\xi$ estimated from the saturation field as a function of the As/Cd flux ratio. Intriguingly, the density of disorder centers ($\sim \xi^{-3}$), shown in Fig. \ref{fig:Fig2}(e), is at least an order of magnitude less than n$_\mathrm{3d}$ for most samples. While n$_\mathrm{3d}$ does not vary significantly with the As/Cd flux ratio, $\xi$ increases steeply as the As/Cd flux ratio is reduced. 

Within the GCDM, MR\% is predicted to be related to disorder by \cite{SongPRB2015}:
\begin{equation}\label{eqn1}
	\mathrm{MR}\% \approx \frac{\mu}{10^4} \, \mu_0 H \frac{\tan \theta_\mathrm{H}}{1+[\tan \theta_\mathrm{H}]^2}
\end{equation}
where MR\% increases with increasing mobility and decreasing $V_0$ when $eV_0 \gtrapprox$ 2 meV. Figure \ref{fig:Fig2}f) demonstrates than our data and Eqn. \ref{eqn1} agree well thus quantitatively verifying the GCDM model.

Evaluation of the quantitative $\xi$ and $V_0$ parameters extracted from MR data allows us to conclude that the main source of disorder influencing MR behavior in Cd$_3$As$_2$ is native point defects. The disorder length scales imply that scattering centers are located every $\sim$ 10-100 nm on average corresponding to point defect densities $\sim 10^{18} - 10^{15}$ cm$^{-3}$, which could reasonably be expected in our epilayers. On the other hand, threading dislocations would be spaced several hundred nanometers or microns apart for the known densities of mid 10$^8$ cm$^2$ in our epilayers \cite{Rice2019}. We also do not expect the As/Cd flux ratio to influence extended defects in the same way as point defects, leading us to this conclusion.

\subsection{Density functional theory}

\begin{figure} 
	\includegraphics[width=0.5\linewidth]{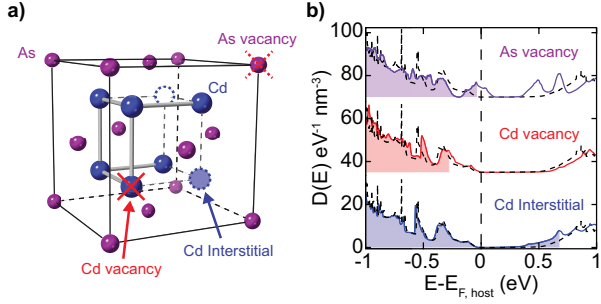}
	\caption{\label{fig:Fig3}(a) Schematic of the simplified Cd$_3$As$_2$ structure derived from the anti-fluorite lattice. Indicated are empty sites on the cation sublattice (blue dashed lines) and the interstitial and vacancy defects. (b) DFT calculated density of states for the V$_\mathrm{As}$, V$_\mathrm{Cd}$, and Cd$_\mathrm{i}$ defects, as function of energy given relative to the Dirac point energy of the defect-free Cd$_3$As$_2$ host.}
\end{figure} 

To understand the role of point defect populations in electron transport in Cd$_3$As$_2$, we examine their behavior with DFT. The 80 atom primitive cell of the Cd$_3$As$_2$ crystal structure (space group $I4_1/acd$) can be considered as being made up of building blocks of anti-fluorite Cd$_4$As$_2$ with an ordered arrangement of missing Cd atoms (empty sites) \cite{MazharInorgChem2014}, as illustrated in Fig. \ref{fig:Fig3}(a). Thus, we expect that Cd interstitials (Cd$_\mathrm{i}$) and vacancies (V$_\mathrm{Cd}$) can form readily by either occupying these empty sites or removing existing Cd atoms, respectively. We additionally consider As vacancies (V$_\mathrm{As}$), which have been tentatively implied as the source of electron doping in early investigations of Cd$_3$As$_2$ \cite{Spitzer1966}. Figure \ref{fig:Fig3}(b) shows the calculated density of states (DOS) of these point defects. The DOS of V$_\mathrm{As}$ shows a localized defect state close to $E_\mathrm{F}$, suggesting that V$_\mathrm{As}$ will cause electron scattering. This behavior is in agreement with scanning tunneling microscopy measurements, which showed that V$_\mathrm{As}$ caused fluctuations in conductance \cite{jeon_landau_2014}. The defect state of the vacancy is half filled, indicating that V$_\mathrm{As}$ has an amphoteric character and is not likely to be the source of free electrons. In contrast, the V$_\mathrm{Cd}$ and Cd$_\mathrm{i}$ defects introduce hole and electron doping, respectively, without significantly changing the DOS profile of the defect-free crystal. We therefore conclude that electron scattering is likely dominated by the localized defect state of V$_\mathrm{As}$, whereas the variation of the electron density n$_\mathrm{3d}$ results from competition between V$_\mathrm{Cd}$ and Cd$_\mathrm{i}$. With increasing As/Cd flux ratio, this balance should reduce the electron doping arising from excess interstitials, a trend that is observed in Fig. \ref{fig:Fig1}(e) for flux ratios above 0.5. A more detailed computational account of defects and doping in Cd$_3$As$_2$ will be presented elsewhere.

\section{Discussion and Conclusion}

We now link the apparent discrepancy between $\xi$ and n$_{3d}$ to the roles of point defects on disorder potentials. If the scattering centers were caused by electron donors and were uniformly distributed, we would expect $n_{3D} \approx \xi^{-3}$. However, $\xi^{-3}<< n_{3D}$ and decays more quickly with increased As/Cd flux ratio (Fig \ref{fig:Fig2}(f)). Furthermore, V$_\mathrm{As}$ is expected to contribute substantially to electron scattering but not to the overall electron concentration. We therefore conclude that V$_\mathrm{As}$ defects are primarily responsible for the disorder length scale, $\xi$, and that an increased density of V$_\mathrm{As}$ leads to decreased electron mobility and MR\%, as those parameters also decrease significantly with lower As flux. This conclusion is supported by a previous bulk crystal study \cite{NarayananPRL2015} that linked V$_\mathrm{As}$ to LMR in a single sample through a temperature-dependent correlation between $\mu$ and MR\%. By systematically controlling the V$_\mathrm{As}$ concentration across a larger set of samples, we are able to experimentally confirm their role in LMR. Both individual defects and clusters of defects can cause the varied potential landscape necessary for LMR in the GCDM framework. Future work should include distinguishing between the two.

In summary, we find that the As/Cd flux ratio controls the concentration of point defects which contribute strongly to scattering. Reduction of the scattering center defect concentration from ~8$\times 10^{16}$  to 3$\times 10^{15}$ cm$^{-3}$ produces an increase in the magnetoresistance from 200\% to 1000\% and an increase in the mobility from 5000 to 18,000 cm$^2$/Vs. We find that the onset of LMR occurs at a lower field in the samples with fewer scattering centers. The As/Cd flux ratio also controls the carrier concentration, which varies from $2.2\times10^{17}$ to $9.7\times10^{16}$ cm$^{-3}$ but is not strongly correlated with electron scattering. Using DFT calculations, we link the highly scattering defects to V$_\mathrm{As}$ and the control of carrier concentration to a combination V$_\mathrm{Cd}$ and Cd$_\mathrm{i}$.

Our combined magnetotransport and thin film growth study reveals the connection between point defects, high electron mobility and large LMR in Cd$_3$As$_2$. The tunable control of point defects made possible by MBE growth allows us to experimentally verify the ability of the GCDM to explain LMR behavior. Since Cd$_3$As$_2$ is a model system, this leads to widely applicable guidelines for engineering electronic properties of topological semimetals. First, the most significant driver of large LMR is high mobility. A correlation between mobility and MR\% was previously observed across a range of TSM materials \cite{Singh2020}. In this paper we control mobility within a single system. Our work reveals that the microscopic mechanism for the change in mobility in Cd$_3$As$_2$ is V$_\mathrm{As}$. More generally, this may be any defect with a large density of states near the Fermi level likely to contribute strongly to scattering. Second, while we find that $V_0$ and $E_\mathrm{F}$ vary with the V$_\mathrm{As}$ concentration, they do not strongly impact the mobility or LMR. Thus, to realize high mobility and large LMR, it is essential to reduce scattering defect densities. The verification of the GCDM as an experimental framework confirms the assertion of the GCDM that topological states are not responsible for LMR. Instead, LMR in Cd$_3$As$_2$ and similar systems is due to diffusive motion of charge carriers around potential barriers created by point defects but crucially, the density of point defects must be low to support high mobilities. This model is relevant when the cyclotron radius is on the order of the spacing between scattering centers. Large Fermi velocities and low carrier concentrations, which are common to many TSMs, allow this to occur at moderate magnetic field, but these lessons are not exclusive to TSMs. These results provide experimentally driven conclusions for the origin of LMR in topological semimetals and pave the way for the use of large LMR in device applications.

Acknowledgments: This work was authored by the National Renewable Energy Laboratory, operated by Alliance for Sustainable Energy, LLC, for the U.S. Department of Energy (DOE) under Contract No. DE-AC3608GO28308. Funding provided by the U.S. Department of Energy Office of Science, Basic Energy Sciences, Physical Behavior of Materials program. The research used High-Performance Computing (HPC) resources of the National Energy Research Scientific Computing Center (NERSC), a DOE-SC user facility located at Lawrence Berkeley National Laboratory, operated under Contract No. DE-AC02-05CH11231. This research also used HPC resources at NREL, sponsored by DOE, Office of Energy Efficiency and Renewable Energy. M.L. and I.A.L. acknowledge partial support from an NSF Award No. DMR-2001376. The views expressed in the article do not necessarily represent the views of the DOE or the U.S. Government. The U.S. Government retains and the publisher, by accepting the article for publication, acknowledges that the U.S. Government retains a nonexclusive, paid-up, irrevocable, worldwide license to publish or reproduce the published form of this work, or allow others to do so, for U.S. Government purposes.

{\tiny {\normalsize {\tiny }}}

%

\setcounter{section}{0}
\setcounter{figure}{0}
\renewcommand{\thefigure}{S\arabic{figure}}

\section{Supplementary Note 1: THIN FILM SYNTHESIS:}

ZnCdTe/ZnTe buffer layers were employed to accommodate the lattice-mismatch between Cd$_3$As$_2$ and the substrate and create a lattice-matched template for Cd$_3$As$_2$ growth as reported in ref. \cite{Rice2019}. The As and Cd fluxes were measured using a beam flux gauge, and the
pressures were corrected using standard sensitivity correction factors.

\section{Supplementary Note 2: X-ray photoemission spectroscopy:}

\begin{figure*}
	\includegraphics[width=1\linewidth]{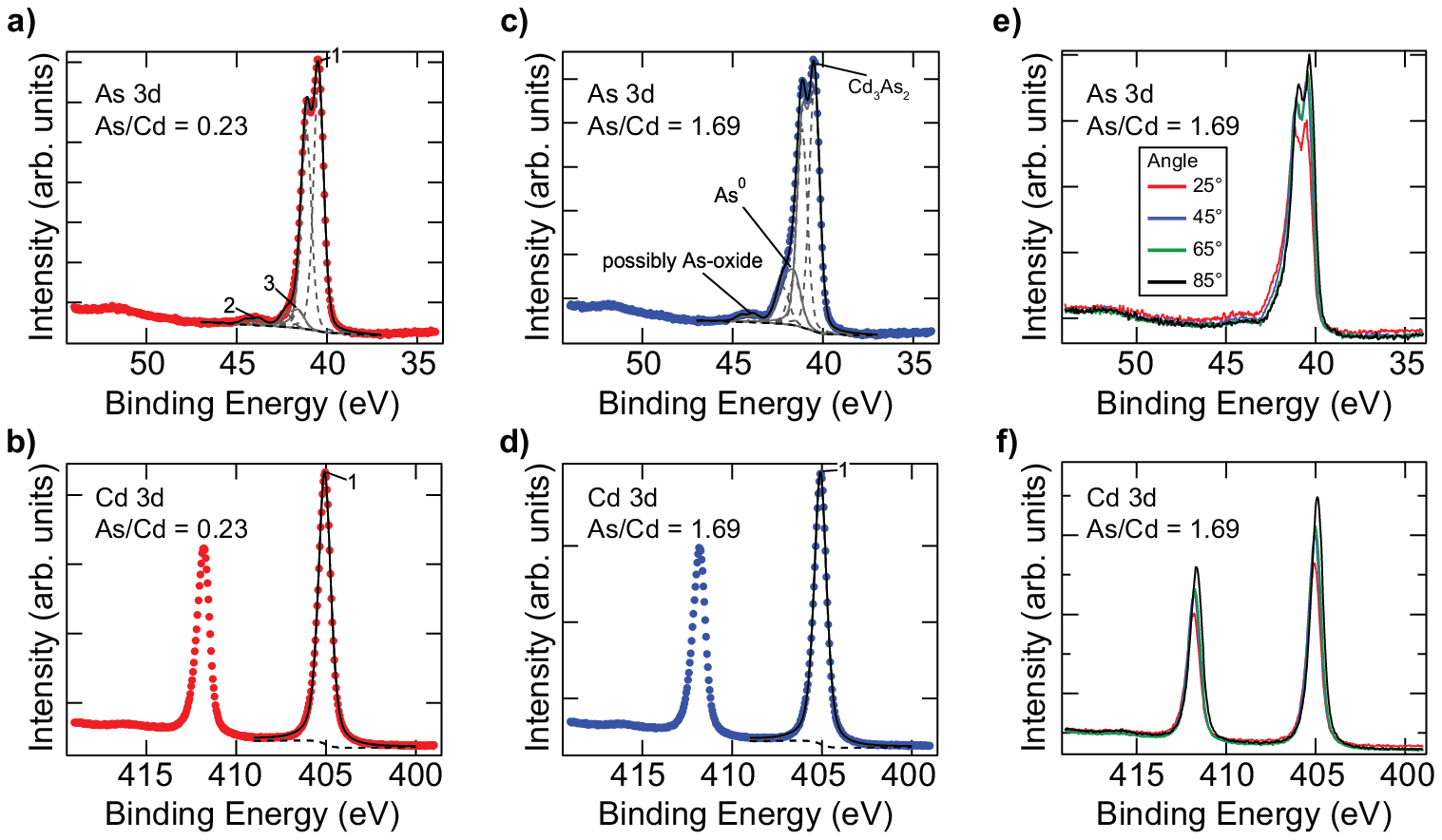}
	\caption{\label{fig:FigXPS} a-d) X-ray photoemission spectra of samples with As/Cd flux ratio of 0.23 (red) and 1.69 (blue) panel showing the As 3d peak (a,c) and Cd 3d peak (b,d). e,f) Angle-resolved XPS measurement of the As/Cd ratio = 1.69 sample demonstrating that the high binding energy shoulder on the As 3d peak is a surface effect. The spectra in c,d) are scaled so that the background intensity is the same for all angles measured, at each angle the Cd 3d and As 3d spectra are scaled by the same factor.}
\end{figure*}

To rule out the possibility that off stoichiometry is driving the observed changes in magnetotransport properties we have performed x-ray photoemission spectroscopy shown in Fig.\ref{fig:FigXPS} a-d). on two samples: one grown with an As/Cd flux ratio of 0.23 (red) and the other grown with a flux ratio of 1.69 (blue). The main difference between the two samples is a small higher binding energy shoulder on the As 3d peak in the As/Cd=1.69 sample. We attribute that peak to excess elemental As on the surface, as confirmed by angle-resolved XPS measurements (Fig. \ref{fig:FigXPS}e,f). They show that this peak is a surface effect as it is more pronounced in measurements with a grazing incidence angle (25$^\circ$), where the surface signal will be stronger than in measurements where the incidence is close to normal (85$^\circ$). Peak fitting, shown in (Fig. \ref{fig:FigXPS}a-d), allows us to isolate the contribution of elemental As and shows that the stoichiometry of the two films are equivalent to within 0.3\%. This conclusively demonstrates that changing the As/Cd flux ratio in the regime studied in this paper does not impact the film stoichiometry.

\section{Supplementary Note 3: Analysis of SdH oscillations:}

\begin{figure*} 
	\includegraphics[width=0.85\linewidth]{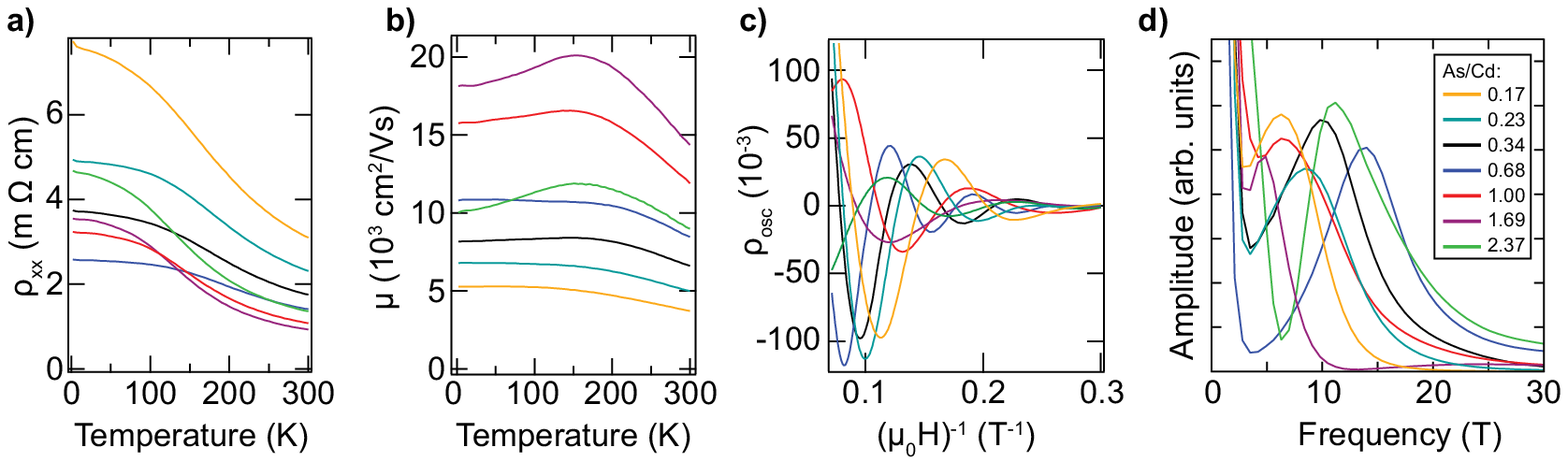}
	\caption{\label{fig:Figrvst} a) Resistivity vs temperature. b) Hall mobility vs temperature.c) $\rho_{osc}$ vs magnetic field. d) Fourier transform of the magnetoresistance showing a single frequency for all samples.}
\end{figure*}

In Fig. \ref{fig:Figrvst}a,b), we show the temperature dependence of the resistivity and mobility. All samples exhibit similar values of the resistivity $\sim 1-8$ m $\Omega$ cm. Fig. \ref{fig:Figrvst} b) shows that the mobility typically increases slightly with lowered temperature and, in some samples, exhibits a peak at intermediate temperatures which has been observed in other Cd$_3$As$_2$ thin films \cite{Galletti21018}. Figure \ref{fig:Figrvst} (c) shows the oscillatory part of the longitudinal MR as a function of inverse field, defined as $\rho_{osc} = (\rho_{xx} - \rho_\mathrm{BG})/\rho_\mathrm{BG}$, where $\rho_\mathrm{BG}$ is a smooth background extracted via a four term polynomial fit. Figure \ref{fig:Figrvst} (d) shows the Fourier transform of the magnetoresistance for all samples demonstrating a single frequency consistent with a single band crossing the Fermi level.

\begin{figure} 
	\includegraphics[width=1\linewidth]{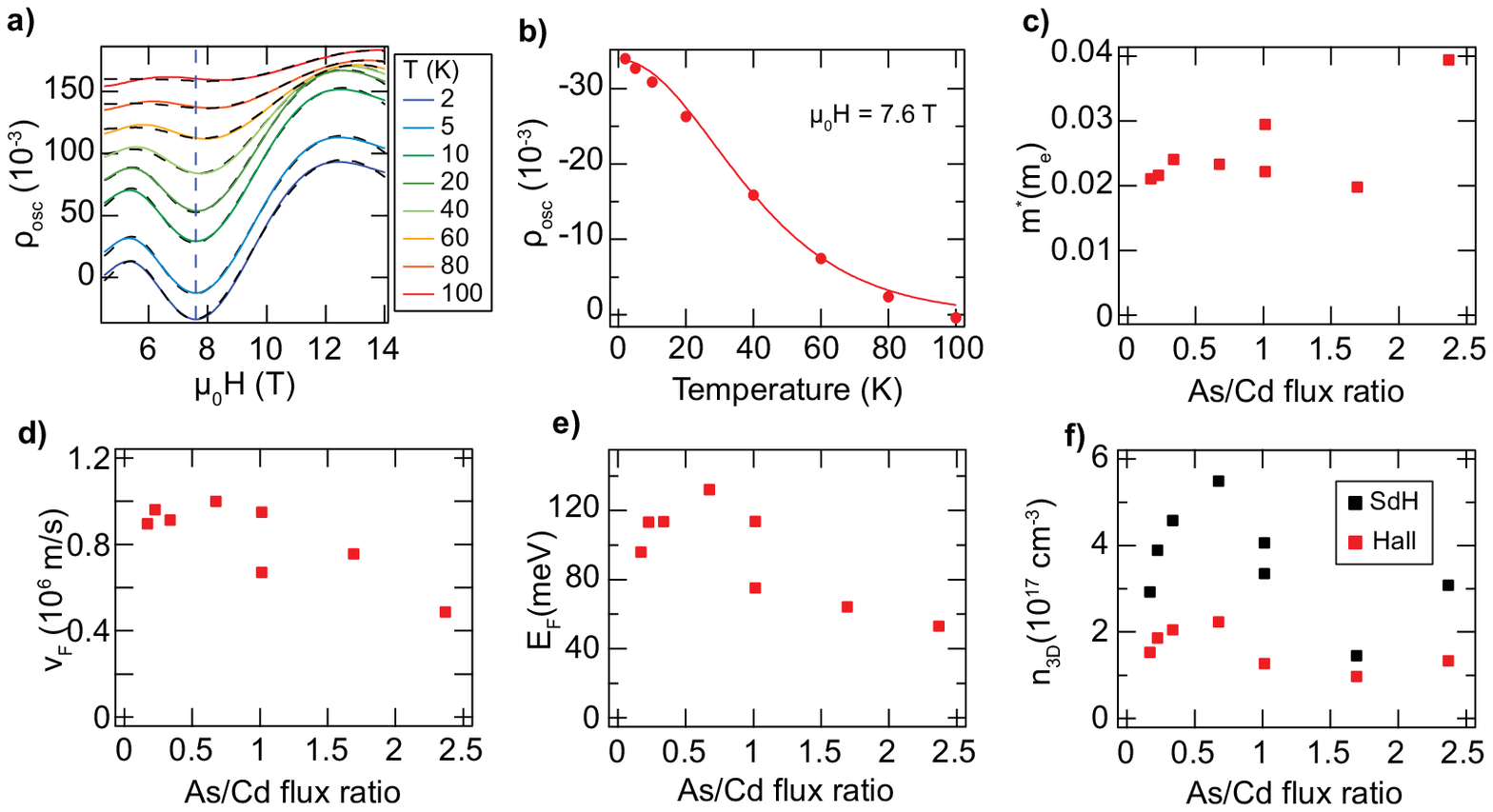}
	\caption{\label{fig:FigLK} a) Example of the oscillatory part of the magnetoresistance extracted using a four term polynomial background, and b) Example of the amplitude of the quantum oscillations as a function of temperature for a sample grown with an As/Cd flux ratio of 1. c-e) parameters extracted by fitting the magnetoresistance data to a standard Lifshiftz-Kosevich function as a function of the As/Cd flux ratio. c) Cyclotron effective mass in units of electron mass, d) Fermi velocity, and e) Fermi Energy. f) carrier concentration determined from the SdH oscillations (black) and low field Hall measurements (red).}
\end{figure}

We use a standard Lifshiftz-Kosevich function of the following form to analyze the SdH oscillations \cite{LiangNmat2015, HePRL2014}:

\begin{equation}\label{eqnLK}
	\rho_{osc} = \frac{\rho_{xx}-\rho_{BG}}{\rho_{BG}} = \left(\frac{\hbar \omega_c}{2 A}\right)^{1/2}\frac{2\pi^2 k_B T/\hbar\omega}{\sinh 2\pi^2 k_B T/\hbar\omega} e^{{-2\pi^2 k_B T_D/\hbar\omega}}\cos\left(\frac{2 \pi F}{B}+\phi\right)
\end{equation}

where $\rho_{osc}$ is the oscillatory part of the magnetoresistance, $\rho_{xx}$ is the resistivity, and $\rho_{BG}$ is a four term polynomial background, $A$ is a constant with dimensions of energy, $F$ is the frequency of oscillations, $k_B$ is the Boltzmann constant, $T$ is the temperature, $\omega_c = eB/m^*$ is the cyclotron frequency and $m^*$ is the cyclotron effective mass. The Dingle temperature is $T_D = \hbar/(2\pi k_B \tau_Q)$ where $\tau_Q$ is the quantum lifetime. For each sample we globally fit constant temperature $\rho_{xx}$ vs $\mu_0 H$ curves at  $T=2,5,10,20,40,60,80,100 K$ to a modified version of equation \ref{eqnLK}:

\begin{equation}
	\rho_{xx}(T=const) \propto \left[\left(\frac{\hbar \omega_c}{2 A}\right)^{1/2} \frac{2\pi^2 k_B T/\hbar\omega}{\sinh 2\pi^2 k_B T/\hbar\omega} e^{{-2\pi^2 k_B T_D/\hbar\omega}}\cos\left(\frac{2 \pi F}{B}+\phi\right)\right] \times\rho_{BG} + \rho_{BG}.
\end{equation}

We then use the resulting $\rho_{osc}$ data (for example Fig. \ref{fig:FigLK} a) to extract the amplitude of a quantum oscillation as a function of temperature at a fixed field (blue dashed line in Fig. \ref{fig:FigLK} a). The amplitude ($\rho_{osc}$) vs T (K) data is then fit to eqn \ref{eqnLK} at a fixed magnetic field as shown in Fig. \ref{fig:FigLK} b). We then iteratively fit until the values converge. The cyclotron effective mass $m^*$ comes directly from this fit (Fig. \ref{fig:FigLK} c). To extract the Fermi velocity for each sample as shown in Fig. \ref{fig:FigLK} d) we make the assumption of a Dirac dispersion $v_F = \hbar k_F/m^*$ \cite{HePRL2014}.  Here $k_F$ is the Fermi crossing  extracted using the Onsager relation between the oscillation frequency and the cross-sectional area of the Fermi surface $F = \frac{\hbar}{2\pi e} \pi k_F^2$. The values of both the cyclotron effective mass and Fermi velocity are similar to bulk single crystals \cite{LiangNmat2015, HePRL2014} and do not change significantly with varied As/Cd flux ratio. To determine the Fermi energy $E_F$ we use the relation for Dirac systems \cite{LiangNmat2015} $m^* = E_F/v_F^2$ (Fig. \ref{eqnLK} e). The carrier concentration is extracted from the $\rho_{osc}$  SdH oscillation period  assuming two identical spherical Fermi surfaces using the relation \cite{ChengNJPhys2016}
\begin{equation}
	n_{3d} = \left(\frac{4}{3} \pi k_F^3 \right) \frac{g_s g_v}{8 \pi^3}
\end{equation}
where $g_s$ is the spin, and $g_v$ is the valley degeneracy. It is apparent that $E_F$ and n$_{3D}$ are correlated as expected for n-type Cd$_3$As$_2$. This carrier concentration (Fig. \ref{eqnLK}(f) (black squared) roughly matches the values from the Hall measurements (red squares).

\section{Supplementary Note 4: Quasiparticle Self-Consistent GW calculations}

To assess our assumptions of the energy bands and Fermi surface made in the SdH
oscillation fitting, we use the Quasiparticle Self-Consistent GW (QSGW) approximation \cite{VanSchilfgaarde2006, Kotani2007}. QSGW is able to reliably estimate bandgaps in most systems, especially weakly correlated ones. It nevertheless systematically overestimates semiconductor bandgaps slightly \cite{VanSchilfgaarde2006}, and for the same reason the velocity at the Dirac point in graphene \cite{VanSchilfgaarde2011}. These overestimates are well understood to originate from RPA approximation to the screened coulomb interaction W, and nearly all of the overestimate can be eliminated with the addition of ladder diagrams to the polarizability \cite{Cunningham2021}. A more efficient, semi-empirical alternative, is to make a hybrid of 80\% QSGW and 20\% LDA. This empirical relation has long been known to yield high-fidelity single-particle levels in semiconductors \cite{Chantis2006,Deguchi2016}. Moreover, it feasible for the large (80 atom) unit cell of Cd$_3$As$_2$, while the more rigorous inclusion of four-point ladder diagrams poses a large challenge. In the results below, we perform a QSGW calculation, making the QSGW
self-energy $\Sigma^0$ self-consistent. When constructing one-particle properties (Fermi surfaces, energy bands) replace $\Sigma^0$ with $0.8(\Sigma^0-V_{xc}^{LDA} )+0.2V_{xc}^{LDA}$ \cite{Deguchi2016}.

\begin{figure*}
	\includegraphics[width=0.67\linewidth]{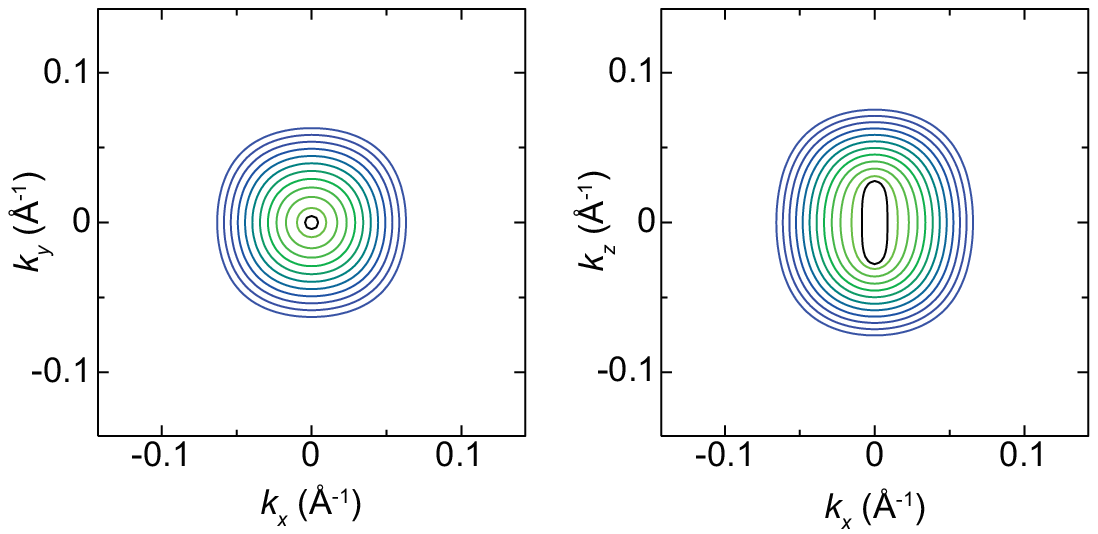}
	\caption{\label{fig:QSGW} a) Constant energy contours in the xy plane passing through $k_z$= (0,0,0.022) \AA$^{-1}$.  The colored contours are in energy increments of 0.025 eV above the Fermi energy.  The black contour is for 0.01 eV.  b) same, for contours in the xz plane passing through ky=0.  All axes are in \AA$^{-1}$ units.}
\end{figure*}

Using this approximation, the Dirac point was found be at (0,0,0.022) in units of the reciprocal lattice vectors, which coincidentally is also (0,0,0.022) \AA$^{-1}$. The Dirac point is somewhat closer to the $\Gamma$ point than what the LDA predicts. Figure \ref{fig:QSGW} shows constant-energy surfaces of Cd$_3$As$_2$ passing through the Dirac point, Fig. \ref{fig:QSGW}a in the plane normal to z with $k_z=\pm0.022$ \AA$^{-1}$, and Fig. \ref{fig:QSGW}b in the plane y=0 passing through $\Gamma$ . The smallest surface (black contour) corresponds to an energy 0.01 eV above the Dirac point (which exactly coincides with the Fermi energy in the undoped case); the remaining contours start at 0.025 eV and increase in uniform increments of 0.025 eV. Fig. \ref{fig:QSGW}a shows that this constant-energy contours are nearly circular in the plane normal to z for chemical potentials less than 0.2 eV. In Fig. \ref{fig:QSGW}b, two minima corresponding to two Dirac points are in evidence at $k_z =\pm 0.022$ \AA$^{-1}$. The bands must be symmetric in $k_z$, so the constant-energy contours reflect two minima that coalesce, making them resemble ellipses at small energy rather than circles. They gradually becoming more circular as the energy increases. As will be shown elsewhere, the energy dispersions stellating from the Dirac point are roughly linear in energy, in the QSGW approximation.

Thus QSGW affirms the two primary assumptions used to relate the chemical potential and carrier concentration: spherical Fermi surface and linear dispersion for small energies. These assumptions are used in our analysis of SdH oscillations presented in supplementary note 3.

\section{Supplementary Note 5: Density functional theory calculations}

It is important to note that GW corrections to the electronic structure of Cd$_3$As$_2$ are necessary for a fully quantitative description of the density of states near the Dirac point. However, we find that the SCAN functional better reproduces the GW band structure than standard DFT functionals, so SCAN is sufficient for demonstrating the qualitative nature of the intrinsic defects and their effect on the electronic states near the Dirac point.


\end{document}